# High-resolution truncated plurigaussian simulations for the characterization of heterogeneous formations

Grégoire Mariethoz[1], Philippe Renard[1], Fabien Cornaton[1], Olivier Jaquet[1,2]

[1]*Centre for Hydrogeology, University of Neuchâtel, 11 Rue Emile Argand, CP 158, CH-2000 Neuchâtel, Switzerland*

[2]*Colenco Power Engineering Ltd, Täfern 26, 5405 Baden, Switzerland*

Tel.: +41 32 718 26 10; E-mail: gregoire.mariethoz@unine.ch

## Abstract

Integrating geological concepts, such as relative positions and proportions of the different lithofacies, is of highest importance in order to render realistic geological patterns. The truncated plurigaussian simulation method provides a way of using both local and conceptual geological information to infer the distributions of the facies and then those of hydraulic parameters. The method (Le Loc'h and Galli 1994) is based on the idea of truncating at least two underlying multi-Gaussian simulations in order to create maps of categorical variable. In this manuscript we show how this technique can be used to assess contaminant migration in highly heterogeneous media. We illustrate its application on the biggest contaminated site of Switzerland. It consists of a contaminant plume located in the lower fresh water Molasse on the western Swiss Plateau. The highly heterogeneous character of this formation calls for efficient stochastic methods in order to characterize transport processes.

## Introduction

The importance of geostatistical characterization of complex geology has increased significantly over the last two decades. To this end, a number of stochastic approaches have been developed. However, the usual continuous and multi-Gaussian methods (Matheron 1965) have shown that they do not allow to model a sufficiently wide range of connectivity patterns for the high (or low) permeable structures (Journel and Alabert 1990; Gómez-Hernández and Wen 1998; Zinn and Harvey 2003; Kerrou et al. 2007; Renard 2007). An alternative is to use a two-step approach in which, first the geological facies are modeled and second, they are populated with heterogeneous hydraulic and transport parameters. This approach is flexible and allows modeling structures at different scales.

When geological facies can be identified from field observations, a large number of methods can be used to simulate such categorical variable (see review in Koltermann and Gorelick 1996; de Marsily et al. 2005). Among the first were the sequential indicator (Journel and Isaaks 1984), the boolean (Haldorsen and Chang 1986) and the truncated Gaussian (Matheron et al. 1987) methods. Sequential indicator simulation is becoming less widely used. It leads to consistency problems, as the generated simulations present multivariate distributions that are implementation-dependent (Emery 2005). Moreover, it fails to accurately represent complex geological structures (e. g. successive deposition and erosion). This is not the case for the Boolean approach that produces realistic geometries (Jussel et al. 1994; Scheibe and Freyberg 1995; Deutsch and Tran 2002). However it is highly parametrized in order to constrain the size, shape, and density of the objects. In the case of large objects, conditioning to data tends to become approximate even if new techniques were recently developed to overcome these issues (Lantuéjoul 2002). The processes of deposition, and erosion can also be simulated in the framework of a mixture of object based and pseudo-genetic algorithm (Webb and Anderson 1996; Cojan et al. 2004). The Markov chain approach (Carle and Fogg 1997) is a powerful alternative which makes use of transition probabilities between the facies. It has been applied to a wide variety of situations (Weissmann and Fogg 1999). More recently, the use of support Vector Machines has been proposed (Kanevski et al. 2002; Wohlberg et al. 2006). The technique can delineate facies using regression techniques, but they do not allow sampling the probability space as they only produce a single facies realization. Multiple-points geostatistics (Strebelle 2002; Caers and Zhang 2004; Feyen and Caers 2006) is a very promising approach. It is based on the concept of single normal equations. The technique offers more flexibility than the Boolean approach while it facilitates conditioning. However, it is still computationally demanding (especially in terms of memory requirements) for large 3D grids with more than 4 facies according to our experience.

In this paper, we investigate the applicability of the truncated plurigaussian method (Le Loc'h and Galli 1994) for high resolution 3D simulation of groundwater flow and transport. The basic idea behind the truncated Gaussian approach is to simulate one or several continuous Gaussian fields and to truncate them in order to produce a categorical variable. To illustrate the idea, let us consider a single Gaussian (continuous) variable and a single truncation: if the simulated Gaussian variable is above the threshold then the point belongs to the facies 1, if it is below the threshold then the point belongs to the other facies. This idea was investigated by Isaaks (1984) and was further developed by Matheron et al. (1987). The main interest of the truncated plurigaussian method is that it allows integrating a geological concept (using a *lithotype rule*) within the framework of a mathematically consistent stochastic model while remaining tractable for large and high resolution 3D grids. It also presents the advantage that conditioning can be achieved in the presence of substantial datasets with acceptable run-times. The technique is for the moment mostly used in the petroleum (Remacre and Zapparolli 2003) and mining industries (Fontaine and Beucher 2006) where it has been widely validated. As far as we know it has not yet been applied to hydrogeology.

To test and demonstrate the applicability of the plurigaussian method, we focus on the case study of the Kölliken contaminated site. This site was chosen because of the existence of an extensive data set (245 borehole logs with geological descriptions, existence of a tunnel, continuous measurements of contamination in observation wells, hydraulic tests, etc).

## Truncated plurigaussian simulations

The mathematical theory underlying the truncated gaussian (Matheron et al. 1987) and the truncated plurigaussian (Le Loc'h and Galli 1994) methods is described in detail in the book of Armstrong et al. (2003) or in the paper of Emery (2007). Therefore, we will emphasize here only the main aspects of the method and present them in an intuitive manner without entering into its mathematical formalism. The principle of the method is to generate two (or more) Gaussian fields, using standard multi-Gaussian techniques, and then to truncate them in order to produce a map of discrete values representing the lithotypes (Le Loc'h and Galli 1994). The statistical inference of the variograms of the underlying Gaussian fields and their conditional generation will be discussed later. Let us first focus first on an illustration of the truncation procedure and on the flexibility that it provides to the modeler. Figure 1 shows two Gaussian random fields: G1 and G2. In that case, G1 has a gaussian variogram model, while G2 has a spherical one and presents an East-West anisotropy. Both fields have a 0 mean and a variance of 1. These two fields are truncated to create 4 lithofacies. Because the relations between the facies can be different depending on the type of geology, the truncation technique has to be flexible. The approach proposed by Le Loc'h and Galli (1994) for that purpose is to define the relations between the facies in a diagram called the *lithotype rule*. Three examples of lithotype rules (A, B, and C) are shown in Figure 1. In these diagrams, the two axes correspond to the values of the underlying multi-Gaussian fields (G1 and G2), and the grey codes correspond to the domain of the different lithotypes. Figure 1 shows the application of the different lithotype rules. With the truncation rule A, G2 is truncated with two thresholds, defining the sand, silt and clay facies. Another threshold has been placed along G1, delimiting the basalt facies. The result (Figure 1a) is that whenever G1 has a low value, the basalt facies is present. At locations with a higher value of G1, sand, silt or clay is present according to the value of G2. Because G1 has a Gaussian covariance, the boundary between the basalt facies and the other facies has a smoother shape than the boundaries between the sands, silts and clasy which is controlled by the spherical model of variogram used to generate G2. The relations and contacts between the facies are imposed by the lithotype rule. In this example, silt can be in contact with all other facies, but sand and clay are not allowed to appear next to each other. On the contrary, basalt can cut all other facies. Note that the surface areas of the different facies in the lithotype rule do not correspond to their respective proportions in the simulation because the underlying continuous variables are not uniformly distributed (they are Gaussian). The control of the proportion in a given simulation requires therefore to compute precisely the value of the threshold (see details in Armstrong et al. 2003). Lithotype rule B has 3 facies that are in a fixed order, silt is a transition facies between sand and clay. One could think that such a result could be obtained by truncating a single Gaussian function. But by looking carefully at the spatial structure of the clay patches (figure 1b), it becomes visible that they are influenced by the spatial structure of both G1 and G2, with some smooth clay patches and other more irregular ones. Lithotype rule C shows that a facies can also be defined by discontinuous zones in the lithotype rule, generating complex effects (Figure 1c).

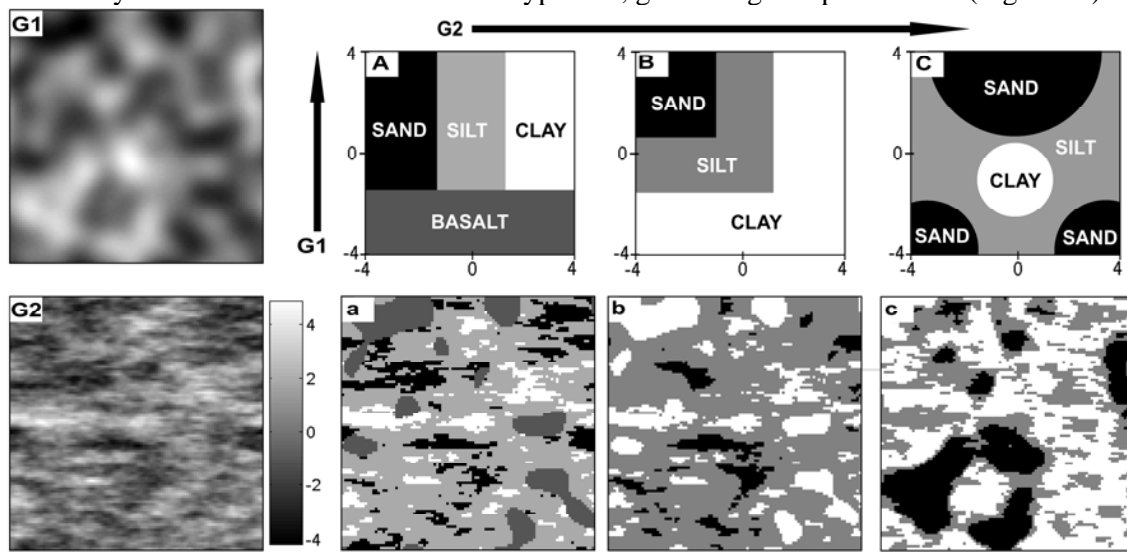

**Figure 1. G1 and G2: Underlying Gaussian fields (100x100 cells). A: Lithotype rule with 4 facies and a: corresponding simulation. B: Lithotype rule with 3 facies, all influenced by both G1 and G2 and b: corresponding simulation. C: Lithotype rule with facies that are defined by discontinuous zones instead of thresholds and c: corresponding simulation. Note that the areas of the different facies in the lithotype rules do not correspond to their respective proportions in the simulation because the underlying continuous variables are not uniformly distributed, but Gaussian.**

The choice of a lithotype rule is therefore a major step of the methodology. In practice, transition probabilities calculated from borehole logs provide good indications on which facies can and cannot be in contact. However, this is not sufficient since it is restricted to the vertical transitions. Therefore the lithotype rule is usually based on both the analysis of the borehole logs and on a geological conceptual model.

To compute precisely the values of the threshold, one needs to define the relative proportions of the different lithofacies that will be simulated. These proportions are estimated by analyzing wells or outcrops data. Furthermore, in most practical cases, these proportions are not constant over the domain, but vary vertically and laterally because of the existence of trends in the geological processes. This non-stationarity is modeled by providing variable proportions over the domain. The lithotype rule is then locally updated by adjusting the values of the thresholds to match the target proportions while preserving the respective positions of the lithofacies.

An important feature of the plurigaussian technique is the inference of the variogram models for the underlying multi-Gaussian fields. Direct adjustment to the experimental variograms is not possible since the only available experimental variograms are the variograms of the indicator functions describing the lithofacies (one per lithofacies, plus all the bivariate combinations) while the two variograms needed for the model are the variograms of the underlying and continuous multi-Gaussian functions. The links between all these variograms are complex functions of the truncation process and of the conditioning to the lithofacies proportions. Therefore, the variogram inference is based on an inverse procedure in which the ranges of the variograms of the multi-Gaussian fields are adjusted iteratively through an inverse procedure. It consists in defining first the type and parameters of the initial variogram models, then this variogram is used to construct a plurigaussian simulation, one can then compute numerically the variograms of the indicators of the facies from the simulated field, and adjust the parameters of the variograms until an acceptable match is obtained between the experimental and the computed variograms (as described in figure 2). This is done automatically using a minimization technique.

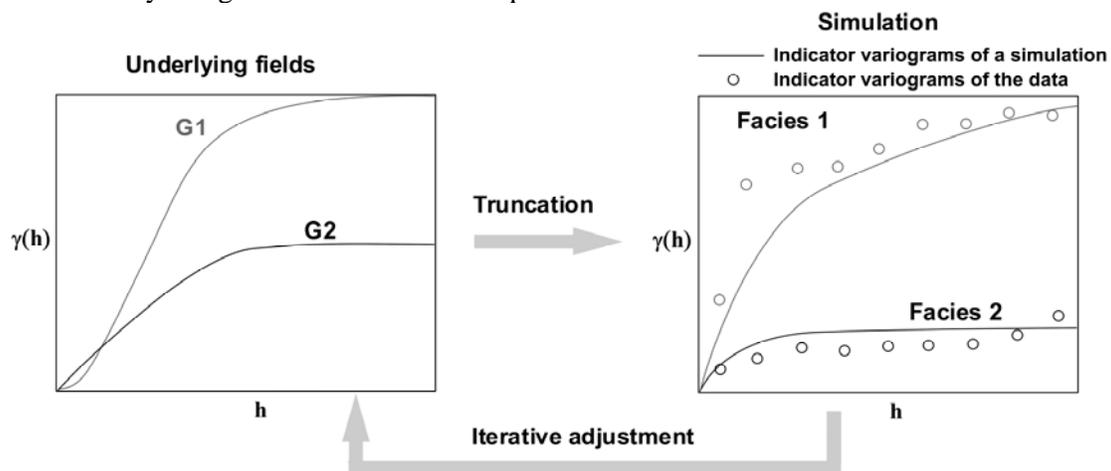

**Figure 2. The variogram models of the underlying Gaussian fields (left) are iteratively adjusted until the indicator variograms of the resulting truncated simulation matches the experimental indicator variograms of the field data (right).**

The last step of the plurigaussian method that needs to be explained is the conditioning to borehole data. Again, because the method is based on the simulation of underlying multi-Gaussian fields and not on the simulation of the indicator variables, the conditioning cannot be direct. Two approaches are possible. The most rigorous is to impose local inequality constraints to the simulation of the multi-Gaussian fields. This can be achieved using the Gibbs sampler (Geman and Geman 1984). This iterative algorithm was adapted to truncated

Gaussian simulations by Freulon and Fouquet (1993) and is described in detail in Armstrong et al. (2003). The principle of the algorithm is to iteratively re-simulate a large number of time the Gaussian field until it reproduces both the structural model (variogram) and the constraints. Starting from an initial simulation, each pixel is resimulated accounting for the previously simulated values and accounting for the constraints. All simulated values that do not satisfy the constraints are rejected by the algorithm. After a certain number of iterations the process converges that has the required properties. A faster approach consists in modifying locally the proportions in order to impose the presence of a unique lithofacies at the conditioning position.

# Application

## *Site description*

The test site is the Kölliken waste landfill in central Switzerland. Between 1978 and 1985, 320'000 tons of special waste materials were buried in this ancient clay quarry. The waste consisted mostly of products of the chemical industry and incineration ashes. Due to economical reasons and improper legislation at that time, no impervious layer was disposed to prevent leakage towards the underlying sandstone units. Moreover, a good drainage system was not implemented. As a result, Kölliken is today the biggest contaminated site in Switzerland. It has been investigated extensively and strong remediation measures have been applied (Abbaspour et al. 1998; Hug 2004; 2005). The main difficulty was the highly heterogeneous character of the site. The geological formations belong to the lower fresh water Molasse of the Swiss plateau. They consist of a succession of sandstones and marls corresponding to a setting of terrestrial deposition with meandering channels (Berger 1985; Sommaruga 1997). Given the amount of non-degradable pollutants, a pump-and-treat approach was first adopted. The remediation project was extended in 2003 by drilling a drainage tunnel along the southern side of the site, collecting the water downstream the landfill on 129 drainage wells at a depth of up to 20 m, combined with an on site water treatment plant. The purpose of this tunnel was to create a piezometric depression stopping further leakage. However, a small part of the plume is already far away downstream and cannot be recovered by pumping. This plume is now advancing in the Molasse formation and may reach the overlying alluvial aquifer. This aquifer supplies drinking water wells, the closest one being 4 kilometres downstream of the landfill. The limit between the alluvial aquifer and Molasse is a smooth but irregular surface of erosion. The motivation to use a high resolution geological model for the Kolliken site is the presence of thin, high permeable and well connected features in the Molasse that one can observe on outcrops. These structures are expected to control most of the fast contaminant migration. The aim of the research being methodological, only the part of the USM that is just below the erosion surface and that contains the part of the contaminant plume not captured by the drainage system has been selected for the model. The model extension (Figure 3) has a rectangular extension parallel to the cardinal directions. The geological data set consists of 245 boreholes logs and 219 measurements taken along the drainage tunnel. 9 cross sections have been made by integrating the data and the geological knowledge (Hug 2005).

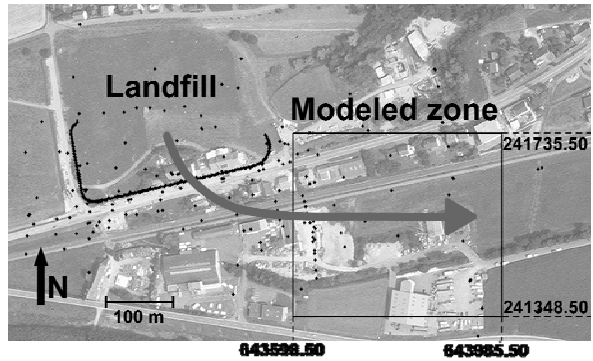

**Figure 3.** Situation of the modelled zone and boreholes location (dots). The line of wells along the drainage tunnel is visible on the southern side of the landfill. Coordinates are in the CH1903 Swiss coordinate system.

## *Conceptual geological model*

A thorough geological analysis of the site is essential in order to build a valid structural model. This is a major step because hard data (such as borehole logs) are usually insufficient to define the type of internal structure of the aquifer.

From a geological standpoint, Molasse is a thick Tertiary sedimentary body created by the detrital filling of a subsidence basin that was caused by the uplift of the Alps. With increasing paleo-distance from the Alps, one can find sediments ranging from very thick alluvial debris fans to deep marine turbidites. The total thickness of the Molasse formation can reach up to 5000 m on its southern side and is thinning up northwards. At the time of deposition, terrestrial debris arrived continuously from the Alps, generating more subsidence. Together with eustatic variations, this led to four different stages of marine and terrestrial deposits (Berger 1985). These four stages are classically described as UMM (first marine stage), USM (first terrestrial stage), OMM (second marine stage) and OSM (second terrestrial stage). The Kölliken site lies within the USM in which the sedimentary structures are a succession of sandstones and marls. Together with paleogeographical and stratigraphic information, the geological interpretation of the area depicts an alluvial plain with meandering rivers. Crevasse splays deposits intersect levees, and the channel belts are wandering through the alluvial plains scattered with marshy spots (Keller et al. 1990; Keller 1992). Very detailed core analyses have identified 42 different facies, some of them highly represented and others very scarcely. Such a level of detail cannot be handled within the plurigaussian simulation framework, and these facies have to be grouped in a set of major lithotypes. Grouping cannot be done according to hydraulic parameters only, as this would lead to group facies with very different types of geometry and connectivity. For example, thin clay layers of lacustrine sediments and thick floodplain deposits may have the same conductivities but the simulation of their spatial distribution must be made separately in order to reproduce these differences and spatial structures because they will have different impacts on flow and transport.

To define which facies need to be modeled, we first have to understand the geology and the architecture of the USM formation. The most active element of such a system is the *river channel* (Figure 5a) moving sideways by erosion and deposition on the outer and inner banks. The point bar, where the slow motion of water allows for deposition of the suspended load and bed load, is mostly made of coarse sediments. A vertical section through a point bar deposit exhibits a gradation from coarser sand at the base to finer at the top (Nichols 1999).

Repeated deposition of sand close to the channel edge leads to the formation of a *levee*, a bank of sediment flanking the channel which is higher than the level of the floodplain. With time, the level of the bottom of the channel can be raised by sedimentation in the channel and the level of water becomes higher than the floodplain level. When the levee breaks, water loaded with sediment is carried out on the floodplain to form a *crevasse splay*, a low cone of sediment formed by water flowing through the breach in the bank and out in the floodplain. These sediments are a heterogeneous mix of coarse debris carried by the river and fine material taken from the levees.

Flooding is not limited to crevasse splays: when the volume of water being supplied to a particular section of the river exceeds the volume which can be contained within the levees, the river floods and over bank flow occurs beyond the limits of the channel. Most of the sediment carried out on the *floodplain* is suspended load which will be mainly clay- and silt-sized debris. As water leaves the channel, it loses velocity very quickly. This drop in velocity triggers the deposition of most of the suspended load as thin sheet over the floodplain. These sheets of sand and silt deposited during floods events are thicker near the channel bank because coarser suspended load is dumped quickly by the flood. In periodically flooded plains, large areas of standing water can develop and persist for years of months. These can be assimilated to small lakes and are represented by finely laminated sediments. On most of the floodplain's area, the marly sediments are in contact with the atmosphere and plants start to colonize this free space. This kind of environment is characterized by thick layers of marls and paleosols.

Though it is not highly tectonized, the USM has endured deformation during the alpine orogenesis. The resulting small scale fracturation is not well documented for the Kölliken site but has been observed in the region. These fractures clearly influence the hydraulic conductivity of the system.

The Molasse formation has then been eroded at the end of the Tertiary period. The final deposition phase was Quaternary alluvial sediments filling the bottom of the valley.

## *Data analysis and grid construction*

An important point that must be clarified before starting the stochastic modeling is that the main directions of continuity of the lithology are generally not horizontal and not constant in space because the sedimentological structures are often deformed by tectonic movements. Therefore, before starting the data analysis, one has to identify what was the horizontal level at the time of deposition. This can be done by identifying in the borehole data base a reference horizon or by investigating the structural data available on the site and interpolating this reference horizon over the whole site even if it does not correspond to a unique lithology. Once this horizon is identified and constructed, the entire domain can be deformed by coordinate transforms in order to restore its initial state as described by Armstrong et al. (2003) (see figure 4).

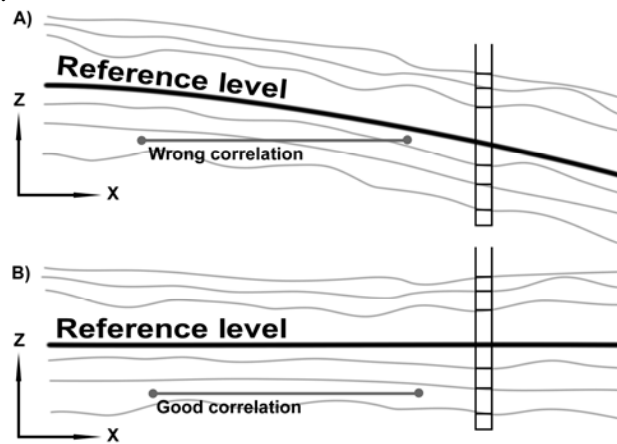

**Figure 4. A) Schematic sedimentary formation with reference level. Correlations inside a single layer are not horizontal. B) The same formation after flattening according to the reference level. Horizontal correlations are rendered possible to compute.**

The geological modeling, including variogram inferences and the generation of conditional plurigaussian simulations, is done in the deformed space. The back transform allows returning to the actual system of coordinates.

For the Kolliken site, a reference level has been constructed by drawing a line following the structures on all the geological sections, ensuring coherence between the sections, and interpolating a 2D surface from the data obtained from this interpretation. The interpolated surface is then used as a reference to modify the vertical component of the coordinate system in order to restore the data to their probable relative position at the time of deposition. A three dimensional regular grid is then built in this system of coordinate. The top of the model corresponds to the erosion surface between the Molasse and the alluvial aquifer (OKF surface). The bottom of the model is an arbitrary flat horizon at 340 m above sea level (about 120 m under the topographic surface). The volume is discretized into 2'984'725 blocks of 3 by 3 by 0.5 m.

## *Truncated plurigaussian simulations*

Based on a preliminary geostatistical study performed by Thakur (2001) and considering the geological conceptual model described above (Figure 5a), five facies have been considered: the channels (RG), the levee (UW), the crevasse splays (DFR), the paleosols in the flood plain (UPS), and the lacustrine deposits (LAK). We decided to model the relations between the facies by the lithotype rule shown in figure 5b. This allows describing the lateral succession from channel to levee. The crevasse splay starts in a breach in the levee and can be in contact with all the other facies. Furthermore, crevasse splays have irregular boundaries as they surge in the alluvial plain and create deposits during sudden events. The first underlying multi-Gaussian function G1 is then modeled with a Gaussian covariance model (resulting in smooth boundaries between the channels and the associated structures) with initial ranges of 250 m in the EW direction, 80 m in the NS direction and 3m vertically. These ranges reflect the lateral and vertical extension of the channels. The second multi-Gaussian function G2 controls the position of the boundary between the crevasse splays and the other facies. On the contrary to G1, the ranges of G2 are longer on the NS direction (150 m) than in the EW direction (100 m) because the two structures are perpendicularly oriented. The initial vertical range is only 1.5 m as crevasse splays form very thin beds. G2 has an exponential variogram, which allows modeling an irregular boundary.

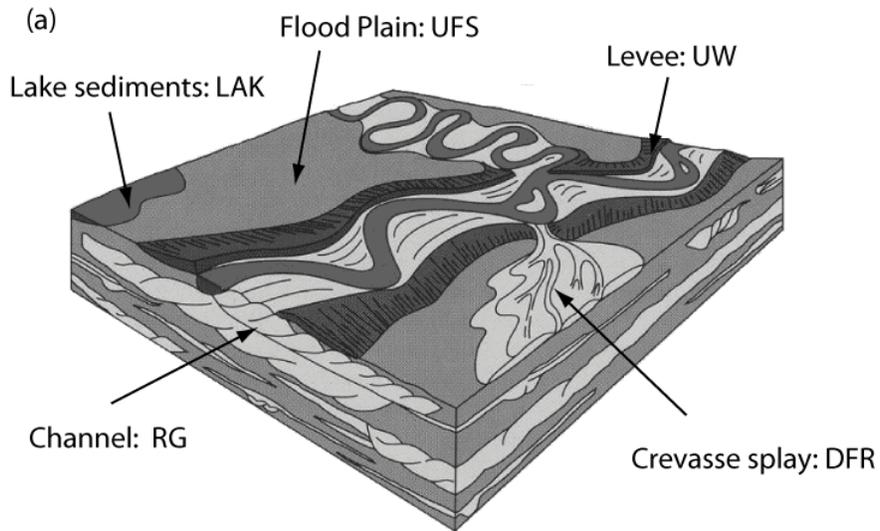

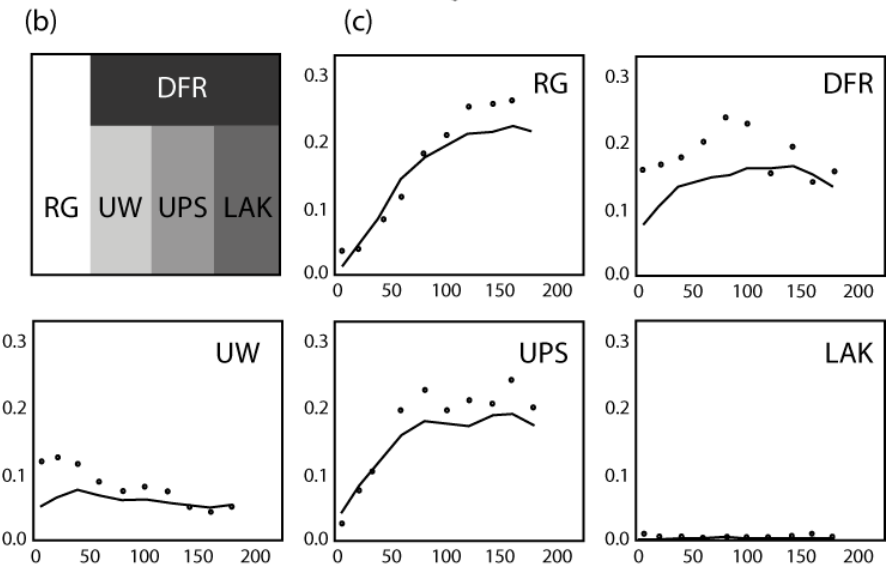

**Figure 5. (a) The conceptual model of the deposition environment and the associated lithotype rule. Modified from (Keller et al. 1990). (b) the lithotype rule, (c) fitted indicator variograms for the 5 lithofacies in the horizontal direction.**

With the prescribed lithotype rule and 464 conditioning data, after having inferred the variograms by inversion (Figure 5b), 300 realizations were generated to attribute facies codes to all grid cells. The position of the grid cells is then back transformed in order to display the results in the actual system of coordinates (figure 6).

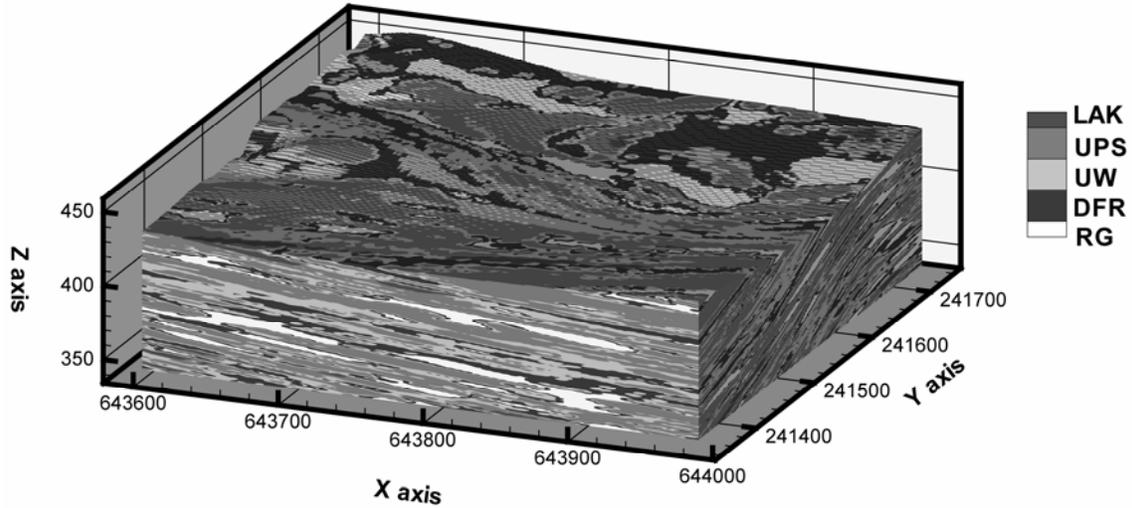

**Figure 6. One example of a conditional realization generated by plurigaussian simulations. The simulation is shown after the back transform to the real coordinate system.**

## *Flow and transport parameters*

Among all the components present in the contaminant plume, Brome was chosen for the simulation as it is a conservative tracer and a clear indicator of the contamination from the landfill. Its migration is modeled on each stochastic realization in transient state with a time-stepping scheme for a period of 15 years with the *Groundwater* finite element code (Cornaton 2006). The flow field is assumed to remain in steady state.

The mean and variance of porosity and hydraulic conductivity for every facies have been measured in the laboratory on small plugs (3 cm high cylinders having a diameter of 3 cm) taken from boreholes cores (Keller et al. 1990; Dolliger 1997). This data set has been collected in the same geological environment but not on the Kölliken site. To distribute the conductivities and permeabilities within the domain, we use this statistical information as well as non-conditional simulation because no data are available at the scale of the model elements on the Kölliken site.

In order to reproduce the correlation between porosity and permeability, porosity was modeled first as the combination of one random multi-Gaussian field for each facies. Note that the variogram used for the porosity simulation could not be inferred from on site data (because these data were not available) and was therefore estimated from geological knowledge by setting realistic ranges. Then the hydraulic conductivities ($K$) were estimated from the porosity with the Hagen-Poiseuille law:

$$K = \frac{n^3 \rho g}{b A_s^2 \mu}, \quad (1)$$

where $n$ is porosity [-], $b$ a formation factor (usually between 10 and 20), $A_s$ the specific contact surface between grains and water [$m^2/m^3$], $\mu$ the water viscosity fixed at 0.0027 [kg/m s], $\rho$ the water density fixed at 999.7 [$kg/m^3$] (for freshwater at 10°C) and $g$ the gravity acceleration, 9.81 [$m^2/s$]. One relation of that type was used for each of the lithofacies. The relation was obtained by adjusting the value of $A_s$ to optimize the fit of the data. The formation factor $b$ could be adjusted as well but because it constitutes a group with $A_s$ it is not possible to identify both of them separately so we decided to keep it fixed and equal to 20.

Furthermore, a white noise is added to represent the natural fluctuations that occur around the mean model. The variance of this noise has been estimated from the variance of the residuals between the measurements and the fitted Hagen-Poiseuille law (Table 1). Two resulting hydraulic conductivity fields are shown in figure 7.

| Facies | Nb of samples | $A_s$ | $\sigma^2 \log_{10}K$ of residuals | Mean $\log_{10}K$ | $\sigma \log_{10}K$ | Mean $n$ | Variance $n$ |
|---|---|---|---|---|---|---|---|
| RG | 35 | 44000 | 1.22 | -5.95 | 1.46 | 0.209 | 0.003 |
| DFR | 23 | 166100 | 1.51 | -7.68 | 1.70 | 0.154 | 0.005 |
| UW | 21 | 700000 | 1.11 | -9.10 | 1.21 | 0.135 | 0.003 |
| UPS/ LAK | 3 | 1200000 | 0.55 | -9.56 | 0.38 | 0.112 | 0.003 |

**Table 1. Summary of the parameters used for the relationship between porosity and hydraulic conductivity. Hydraulic conductivity is expressed in [m/s].**

In terms of transport parameters, the main dispersive process at the scale of the model is believed to be represented by the heterogeneity of the geological structure, therefore we set the longitudinal dispersivity coefficient to a value of 10 m and the transversal dispersivity to 1m.

## *Iintial and boundary conditions*

The initial distribution of the Brome concentrations is estimated by kriging 36 values measured during spring 2005 and ranging from 0.02 to 0.24 mg/l. This procedure is not optimal as the kriged field is not conditional on the geology and not constrained by the physics of solute transport. For example, it is highly probable that the contamination has not entered the low permeability formations and this is not accounted for. Furthermore, the uncertainty on this initial field is not evaluated while the variograms resulting from such sparse dataset are uncertain. All these limitations would have to be overcome in the case of an application with real practical implications.

As the limits of the model do not coincide with hydrogeological limits, it is not possible to prescribe boundary conditions corresponding to real physical boundaries. Nevertheless, previous regional models indicate that a bidirectional flow takes place on the vertical direction. This flow is driven by two nested flow systems. The first is local. It is caused by rainwater infiltrating on the escarpments and emerging in the bottom of the Kölliken valley in the modelled zone. It causes locally an upward flow component. The other system takes place at a bigger scale. A deep karstified calcareous bank underlying the USM drains the area and causes a general downward flow component. A water divide surface at depth within the Molasse separates the two systems. To represent this complex hydrogeological situation, it was chosen to impose head conditions on all sides of the model and to use a series of 1D Hermite polynomial interpolations along the vertical boundaries to force bidirectional flow. Again in order to limit the complexity of the case study, the transient variations of the head at the boundaries of the domain or the uncertainty on the boundary conditions are not accounted for.

The boundary conditions of the transport problem are kept very simple. The drainage tunnel, that lies downstream of the landfill and which was built in 2001, creates a piezometric depression capturing now all the leaking contamination. The consequence is that no new contaminant is arriving in the modelled area and this is why a zero concentration is prescribed to all inflowing zones of the system.

## *Model calibration*

When simulating transport with the values of porosity and hydraulic conductivity described earlier, the plume migration was much slower than observed in the field. This difference is attributed mainly to the presence of small scale fracturation which is not accounted for in the laboratory measurements on small plugs (sampling bias). To reproduce the mean velocity of the plume, the hydraulic conductivity needs to be increased. This observation is in agreement with previous description of the so-called scale effect in permeability (Kiraly 1975; Clauser 1992; Schulze-Makuch and Cherkauer 1998; Zlotnik et al. 2000). We note that in addition to the sampling bias, there is an upscaling effect (Renard and de Marsily 1997; Neuman and Di Federico 2003) because the rock sample size is smaller than the grid blocks used in the model. This effect tends to reduce the variance and increase slightly the geometric mean of the distribution of the hydraulic conductivities in 3D, if we assume a classical multi-Gaussian model. This is however not sufficient to explain the discrepancy between the observed and modeled plume velocity. Therefore, we interpret this effect as due to the presence of small scale fractures within the USM formation which increase significantly the conductivity, but not the porosity. If all facies were made of consolidated rock, the fractures would have a uniformly distributed aperture. Therefore, the hydraulic conductivity of the fractures could just be added to the one of the matrix. Here, we made the hypothesis that the fractures are more open in consolidated sandstone than in clay or marls and, therefore, this effect increases the permeability differences between these facies. This can be rendered by multiplying all permeabilities by a given factor which has been adjusted by trial and error. A factor of 10 gave the best match. Note that an alternative approach could have been to reduce the effective porosity, however this would require that either there is a bias in the sampled porosity, or that the upscaling should reduce in average the porosity. This last assumption is incorrect since porosity is an additive variable. The first explanation cannot be completely rejected but if there is a bias, it may well be in the opposite direction because of the observed presence of the small scale fractures that tends to increase the overall porosity and not to decrease it. Something that cannot be rejected is the fact that there could be some slight geological difference between the Molasse in Kölliken and in the other locations were the measurements have been taken. Diagenesis may for example have been stronger in Kölliken. But again, we do not have sufficient data to support this hypothesis, and therefore we decided to use the simplest explanation compatible with our observations. Two of the resulting hydraulic conductivity fields are displayed in Figure 7.

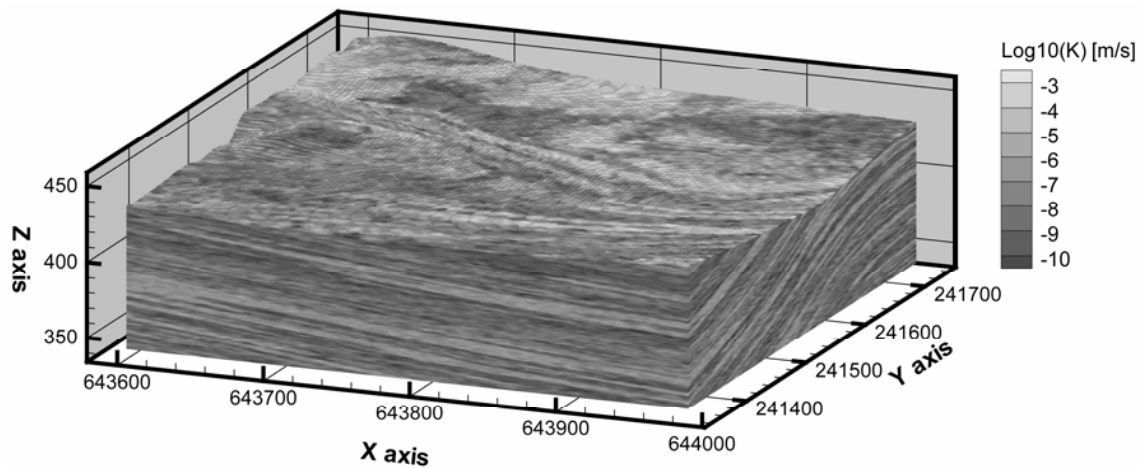

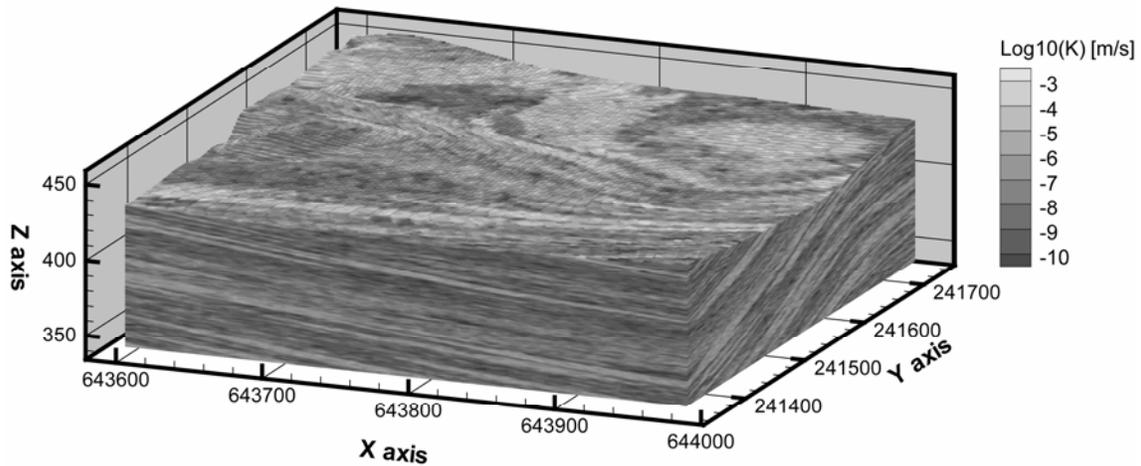

**Figure 7. The hydraulic conductivity field in log scale for two different realizations. While the general structure is similar, both realizations are clearly different, and lead to different contamination results.**

## *Simulation results*

The result of this procedure is a dataset of 300 realizations of concentration fields varying as a function of space and time. Overall, the plume is following the flow field and exits progressively the studied zone. The amount of contaminant leaving the area can thus be compared to the falling limb of a breakthrough curve. Figure 8 shows two vertical cross-sections through the domain for a particular realization at three time steps. From these raw results different statistical measures can be estimated, such as maps of mean concentration, probability maps (e.g. probability of having a concentration higher than a given threshold during a certain period of time) or a statistical distribution of global contamination fluxes.

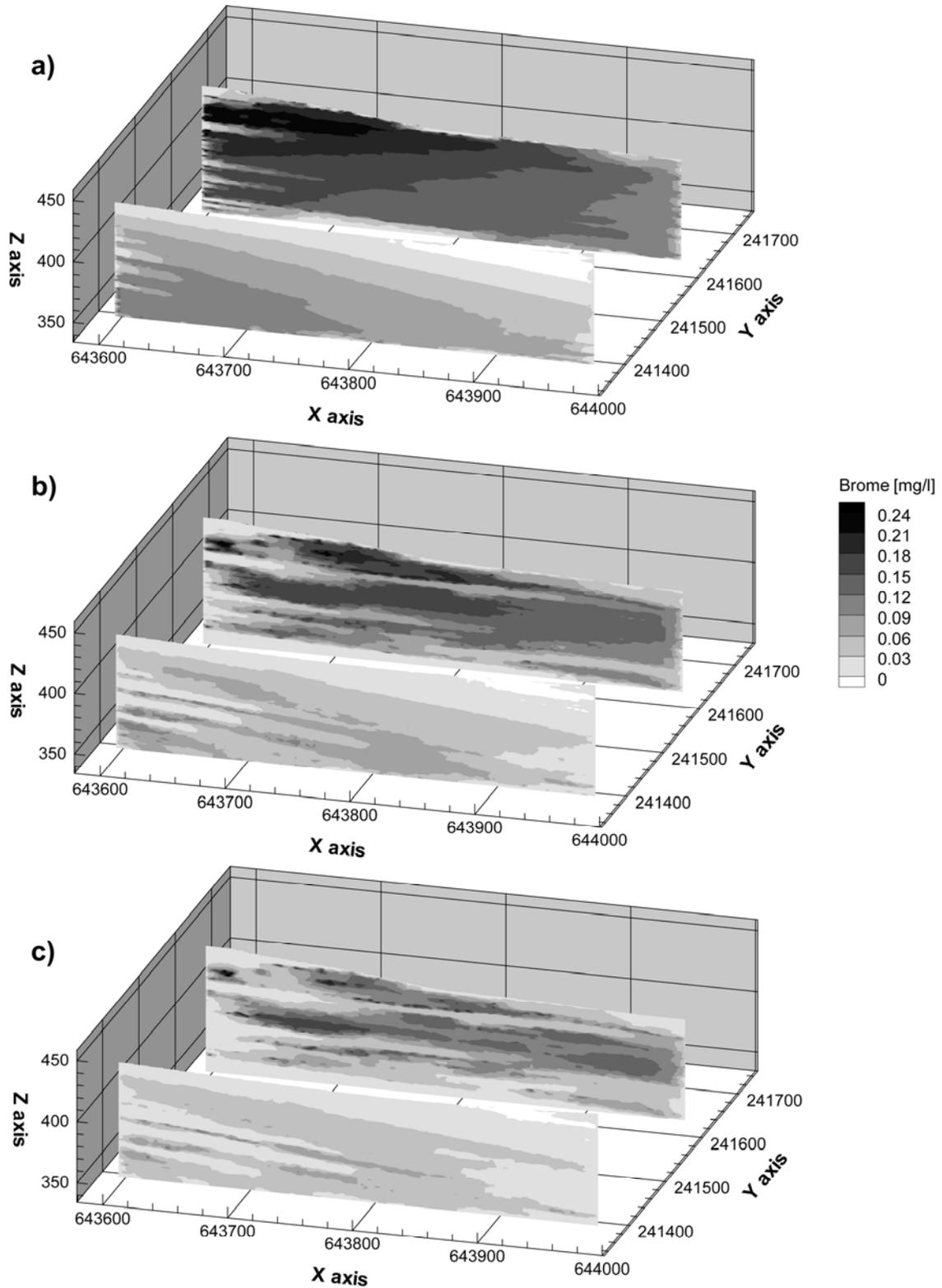

**Figure 8. a) The plume evolution after 0.5 years. b) 5 years. c) 10 years.**

Here, we present only the breakthrough curves, i.e. the history of the contaminant flux through the surface bounding the model on its eastern side (Figure 9a). After a very fast drop of the flux during the first year, the simulations show an important variability with values ranging from 2 to 5 kg/y. These fluxes slowly decrease with time. For the sake of comparison, the same calculations are made on a naïve homogeneous model with a constant homogeneous equivalent hydraulic conductivity. The value for this hydraulic conductivity has been estimated using the Landau-Lifshitz-Matheron conjecture for 3D isotropic media (Dagan 1993; Noetinger 1994; Renard and de Marsily 1997):

$$K_{eq} = \langle k^{1/3} \rangle^3 \qquad (2)$$

where the brackets <> represent the average of the values of the local hydraulic conductivities $k$. We made this calculation on two data sets: the ensemble of all the hydraulic conductivities estimated from slug tests conducted on site, and the ensemble of our simulated (and calibrated) hydraulic conductivity fields. In the first case, we obtain $K_{eq}=3.4\ 10^{-6}$ m/s and in the second case, we obtain $K_{eq}=1.4\ 10^{-6}$ m/s. These two numbers are in good agreement and confirm that the factor of 10 used for the calibration of the hydraulic conductivities is reasonable. For the homogeneous model, we take the first value estimated from the slug tests as this is the value that one may have taken if the geological model would not have been constructed. The porosity is constant and equal to the arithmetic average of the local porosities: $n=0.17$. Results show that the homogeneous model underestimates the fluxes by a factor ranging from 1 to more than 2 orders of magnitude (Figure 9). Indeed, the homogeneous model does not account for the preferential flow paths formed by the channels and, as a result, the progression of the plume is much slower than in the heterogeneous case.

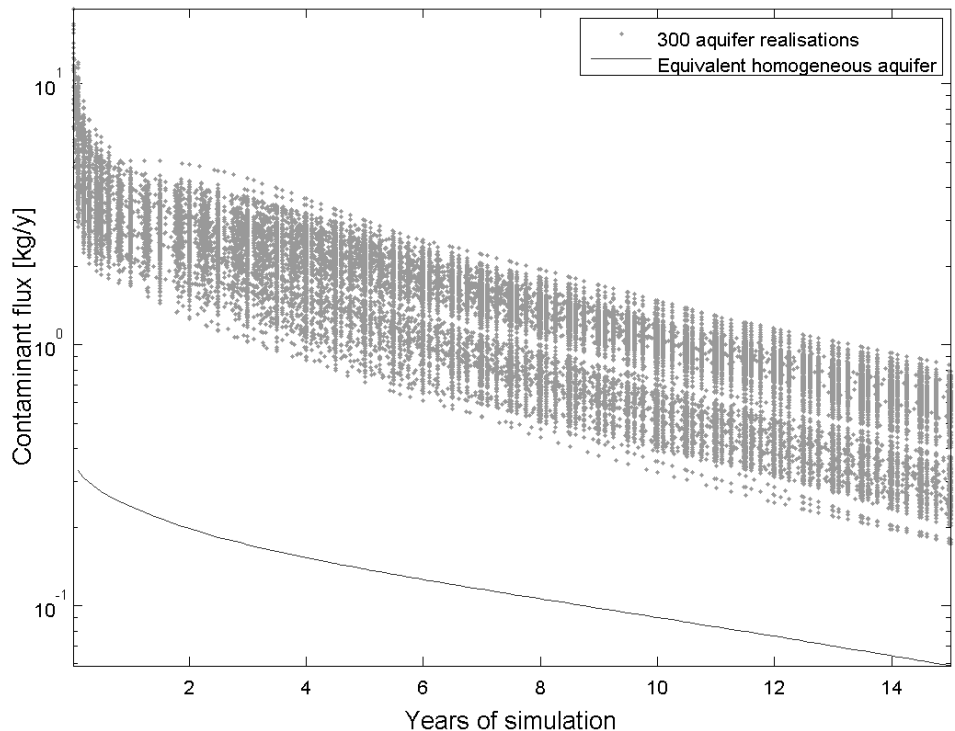

**Figure 9. Breakthrough on the eastern surface for 300 heterogeneous aquifer realizations.**

## Conclusion and discussion

Geostatistical models often suffer from a lack of geological realism due to restrictive assumptions such as for e.g. multi-Gaussianity and maximum entropy. The plurigaussian simulation technique constitutes an interesting tool to overcome these limits. It proposes an efficient way for integrating a geological concept in the stochastic model. As expected, high resolution truncated plurigaussian simulations can model highly connected and highly permeable geological structures that control fast flow and solute transport. On the opposite, using a homogeneous equivalent model leads to a significant underestimation of the contamination fluxes as compared to the stochastic model. This highlights the need to account for those structures. As mentioned in the introduction, several methods can be used to model realistic lithofacies architecture. They all have some limitations and advantages.

The main difficulty in applying the plurigaussian technique is the inference of the variogram models for the underlying multi-Gaussian functions. The process is iterative and depends on the choice of an initial set of variogram parameters that may be difficult to identify. The procedure may fall into local optimums depending on the initial guess. In that case, it may be difficult to justify the use of a given variogram model. Robust optimization techniques such as genetic algorithms could be used to improve that step. In the meantime, we think that an important criterion to accept a variogram model is its coherence with the geological interpretation of the site. The other limitation of the method that we identified during that work is that the lithotype rules are not defined with respect to given directions. Hence, it is not possible to impose to impose, for example, that the levees are always on the side of the channel and not on the top. When a contact is defined in the lithotype rule, it may occur in all directions. This is however compensated by the possibility of imposing non stationarity on the proportions and changing for example the proportions of the different facies with depth.

Overall, the paper illustrates that the truncated plurigaussian simulation technique offers a wide range of interesting features such as the possibility of integrating spatial relationships between the facies, proportion trends, and conditioning to local data. All these caracteristics makes the technique appealing for practical applications in which a conceptual geological model is available. In our view, a particularly interesting aspect of the method is that the definition of the lithotype rule requires a close collaboration between the modeller and the field geologist. It forces a discussion between those two communities that work too often independently and do not share their expertise. Furthermore, the method has a solid mathematical basis with a well defined underlying statistical model and has been validated in different applications in the petroleum and mining industries.

## Acknowledgments

We thank the Swiss National Science Foundation for funding this work (grant PP002-1065557), Olaf Haag of SMDK AG for kindly providing the data that made this study possible, Rudolf Kocher and Rainer Hug at CSD AG Aarau for their cooperation in the field and their kind support. We also thank Andres Alcolea, Fritz Schlunegger, Albert Matter, and Jean-Pierre Berger, as well as two anonymous reviewers, Mary P. Anderson, and Alberto Guadagnini for their constructive comments.

## References

Abbaspour, K.C., Schulin, R., Genuchen, M.T.v., and Schläppi, E. 1998. Procedures for uncertainty analyses applied to a landfill leachate plume. *Ground Water* **36:** 874-883.

Armstrong, M., Galli, A.G., Loc'h, G.L., Geoffroy, F., and Eschard, R. 2003. *Plurigaussian Simulations in Geosciences*, Berlin Heidelberg New York, pp. 149.

Berger, J.-P. 1985. La transgression de la molasse marine supérieure (OMM) en Suisse occidentale. In *Institut de géologie*, pp. 208. Université de Fribourg, Fribourg.

Caers, J., and Zhang, T. 2004. Multiple-point geostatistics: a quantitative vehicle for integrating geologic analogs into multiple reservoir models. In *Integration of outcrop and modern analog data in reservoir models*, pp. 383-394. AAPG memoir 80.

Carle, S.F., and Fogg, G.E. 1997. Modeling spatial variability with one and multi-dimensional continuous Markov chains. *Mathematical Geology* **7:** 891-918.

Clauser, C. 1992. Permeability of Crystalline Rocks. *Eos, Transactions, American Geophysical Union* **73:** 233-240.

Cojan, I., Fouche, O., Lopez, S., and Rivoirard, J. 2004. Process-based reservoir modelling in the example of meandering channel. In *Geostatistics Banff*. (eds. O. Leuangthong, and C.V. Deutch), pp. 611-619. Springer.

Cornaton, F. 2006. *GroundWater, A 3-D Ground water flow and transport finite element simulator*, pp. 220.

Dagan, G. 1993. High-order correction of effective permeability of heterogeneous isotropic formations of lognormal conductivity distribution. *Transport in Porous Media* **12:** 279-290.

de Marsily, G., Delay, F., Gonçalvès, J., Renard, P., Teles, V., and Violette, S. 2005. Dealing with spatial heterogeneity. *Hydrogeology J* **13:** 161-183.

Deutsch, C.V., and Tran, T.T. 2002. FLUVSIM: a program for object-based stochastic modeling of fluvial depositional systems. *Computers and Geosciences* **28:** 525-535.

Dolliger, J. 1997. *Geologie und Hydrogeologie der Unteren Süsswassermolasse im SBB-Grauholztunnel bei Bern*. Geol. Berichte, pp. 44.

Emery, X. 2005. Properties and limitations of sequential indicator simulation. *Stochastic Environmental Research and Risk Assessment* **6:** 414-424.

Emery, X. 2007. Simulation of geological domains using the plurigaussian model: New developments and computer programs. *Computers & Geosciences* **33:** 1189-1201.

Feyen, L., and Caers, J. 2006. Quantifying geological uncertainty for flow and transport modelling in multi-modal heterogenesous formations. *Advances in Water Resources* **29:** 912-929.

Fontaine, L., and Beucher, H. 2006. Simulation of the Muyumkum uranium roll front deposit by using truncated plurigaussian method. In *6th international mining geology conference*, Darwin.

Freulon, X., and Fouquet, C. 1993. Conditioning a Gaussian model with inequalities. In *Geostat Troia '92*. (ed. Soares), pp. 201-212. Kluwer, Dordrecht.

Geman, S., and Geman, D. 1984. Stochastic relaxation, Gibbs distribution and the Bayesian restoration of images. *IEE Trans. Pattern Anal. and Mach Intel.* **6:** 721-741.

Gómez-Hernández, J.J., and Wen, X.-H. 1998. To be or not to be multi-gaussian? A reflection on stochastic hydrogeology. *Adv Water Res* **21:** 47-61.

Haldorsen, H.H., and Chang, D.M. 1986. Notes on stochastic shales from outcrop to simulation models. In *Reservoir characterization*. (eds. L.W. Lake, and H.B. Carrol), pp. 152-167. Academic, New York.

Hug, R. 2004. Entwicklung eines Sanierungskonzeptes für die Spitze der Schadstofffahne der Sondermülldeponie Kölliken (AG), pp. 133. Neuchâtel, CHYN.

Hug, R. 2005. Hydrogeologische Untersuchungen im Abstrom der Sondermülldeponie Kölliken. *Bull. angew. Geol.* **10:** 65.

Isaaks, E. 1984. Indicator simulation: Application to the simulation of a high grade uranium mineralization. In *Geostatistics for Natural Resources Characterization, Part 2*. (ed. G.V.e. al.), pp. 1057-1069. D. Reidel Publishing Company.

Journel, A.G., and Alabert, F.G. 1990. New method for reservoir mapping. *Journal of Petrolem Technology* **42:** 212-218.

Journel, A.G., and Isaaks, E.H. 1984. Conditional indicator simulation: application to a Saskatchewan deposit. *Math Geol* **16:** 685-718.
Jussel, P., Stauffer, F., and Dracos, T. 1994. Transport modeling in heterogeneous aquifer: 1. Statistical description and numerical generation of gravel deposits. *Water Resour Res* **30:** 1803-1817.
Kanevski, M., Pozdnukhov, A., Canu, S., and Maignan, M. 2002. Advanced Spatial Data Analysis and Modelling with Support Vector Machines. *International Journal on Fuzzy Systems***:** 606-615.
Keller, B. 1992. Hydrogeologie des schweitzerischen Molasse-Beckens: Aktueller Wissensstand und weiterführende Betrachtungen. *Eclogae geol. helv.* **85:** 611-651.
Keller, B., Bläsi, H.-R., Platt, N.H., Mozley, P.S., and Matter, A. 1990. *Technischer Bericht 90-41, Sedimentaere architektur der distalen unteren Süsswassermolasse und ihre beziehung zur diagenese und den Petrophysik. Eigenschaften am Beispiel der Bohrungen Langenthal*. NAGRA, Baden, pp. 100.
Kerrou, J., Renard, P., Hendricks-Franssen, H.-J., and Lunati, I. 2007. Issues in characterizing heterogeneity and connectivity in non-multi-Gaussian media. *Advances in Water Resources* **Manuscript submitted**.
Kiraly, L. 1975. Rapport sur l'état actuel des connaissances dans le domaine des caractères physiques des roches karstiques [Report on the present knowledge of the physical characters of karstic rocks]. In *Hydrogeology of Karstic Terrains*. (eds. A. Burger, and L. Dubertret), pp. 53-67. International Association of Hydrogeologists, International Union of Geological Sciences, Paris, France.
Koltermann, C.E., and Gorelick, S.M. 1996. Heterogeneity in sedimentary deposits: A review of structure-imitating, process-imitating, and descriptive approaches. *Water Resour Res* **32:** 2617-2658.
Lantuéjoul, C. 2002. *Geostatistical simulation. Models and algorithms.*, Berlin, Heidelberg, New York.
Le Loc'h, G., and Galli, A.G. 1994. Improvement in the truncated Gaussian method: combining several Gaussian functions. In *Ecmor 4, 4th European Conference on the Mathematics of Oil Recovery*, Roros, Norway.
Matheron, G. 1965. *Les variables régionalisées et leur estimation*. Masson, Paris.
Matheron, G., Beucher, H., Galli, A., Guérillot, D., and Ravenne, C. 1987. Conditional simulation of the geometry of fluvio-deltaic reservoirs. In *62nd Annual Technical Conference and Exhibition of the Society of petroleum Engineers*, pp. 591-599. SPE Paper 16753, Dallas.
Neuman, S.P., and Di Federico, V. 2003. Multifaceted nature of hydrogeologic scaling and its interpretation. *Reviews of geophysics* **42:** 1014.
Nichols, G. 1999. *Sedimentology and Stratigraphy*, pp. 355.
Noetinger, B. 1994. The effective permeability of a heterogeneous porous medium. *Transport in Porous Media* **15:** 99-127.
Remacre, A.Z., and Zapparolli, L.H. 2003. Application of the plurigaussian simulation technique in reproducing lithofacies with double anisotropy. In *Revista Brasileira de Geociências* pp. 6.
Renard, P. 2007. Stochastic hydrogeology: what professionals really need? *Ground Water* **45:** 531-541.
Renard, P., and de Marsily, G. 1997. Calculating equivalent permeability: A review. *Adv Water Res* **20:** 253-278.
Scheibe, T.D., and Freyberg, D.L. 1995. Use of sedimentological information for geometric simulation of natural porous media structure. *Water Resour Res* **31:** 3259-3270.

Schulze-Makuch, D., and Cherkauer, D. 1998. Variations in hydraulic conductivity with scale of measurement durong aquifer tests in heterogeneous, porous carbonate rocks. *Hydrogeology Journal* **6:** 204-215.

Sommaruga, A. 1997. *Geology of the central Jura and the molasse basin*, pp. 176.

Strebelle, S. 2002. Conditional simulation of complex geological structures using multiple point statistics. *Mathematical Geology* **34:** 1-22.

Thakur, R.K. 2001. Geostatistical simulations for 3D geological modelling of an industrial waste landfill, pp. 61. Ecole des Mines de Paris, Fontainebleau (France).

Webb, E.K., and Anderson, M.P. 1996. Simulation of preferential flow in three-dimensional, heterogeneous conductivity fields with realistic internal architecture. *wrr* **32:** 533-545.

Weissmann, G.S., and Fogg, G.E. 1999. Multi-scale alluvial fan heterogeneity modeled with transition probability geostatistics in a sequence stratigraphic framework. *Journal of Hydrology* 48-65.

Wohlberg, B., Tartakovski, D.M., and Guadagnini, A. 2006. Subsurface characterization with support vector machines. *IEEE Transactions on Geoscience and Remote Sensing* **44:** 47-57.

Zinn, B., and Harvey, C.F. 2003. When good statistical models of aquifer heterogeneity go bad: A comparison of flow, dispersion, and mass transfer in connected and multivariate Gaussian hydraulic conductivity fields. *Water Resour Res* **39:** 1051.

Zlotnik, V.A., Zurbuchen, B.R., Ptak, T., and Teutsch, G. 2000. Support volume and scale effect in hydraulic conductivity: experimental aspects. In *Theory, modelling, and field investigations in hydrogeology: A special volume in homor of Shlomo P. Neuman's 60th birthday*. (ed. D.Z.a.L.L. Winter), pp. 215-231. Geological Society of America, Boulder, Colorado.